\def\*{\vskip0.5cm}
\overfullrule=0pt
\newcount\mgnf  
\mgnf=0

\ifnum\mgnf=0
\def\openone{\leavevmode\hbox{\ninerm 1\kern-3.3pt\tenrm1}}%
\def\*{\vglue0.3truecm}\fi
\ifnum\mgnf=1
\def\openone{\leavevmode\hbox{\ninerm 1\kern-3.63pt\tenrm1}}%
\def\*{\vglue0.5truecm}\fi
\ifnum\mgnf=0
\magnification=\magstep0
\hsize=15truecm\vsize=23truecm
\parindent=4.pt\baselineskip=0.45cm
\font\titolo=cmbx12
\font\titolone=cmbx10 scaled\magstep 2
\font\cs=cmcsc10

\font\msytw=msbm10
\font\msytww=msbm8

\font\indbf=cmbx10 scaled\magstep1

\fi
\ifnum\mgnf=2
   \magnification=\magstep0\hoffset=0.truecm
   \hsize=15truecm\vsize=24.truecm
   \baselineskip=18truept plus0.1pt minus0.1pt \parindent=0.9truecm
   \lineskip=0.5truecm\lineskiplimit=0.1pt      \parskip=0.1pt plus1pt
\font\titolo=cmbx12 scaled\magstep 1
\font\titolone=cmbx10 scaled\magstep 3
\font\cs=cmcsc10 scaled\magstep 1
 1

\font\msytw=msbm10 scaled\magstep1
\font\msytww=msbm8 scaled\magstep1

\font\indbf=cmbx10 scaled\magstep2
\fi

\global\newcount\numsec\global\newcount\numapp
\global\newcount\numfor\global\newcount\numfig\global\newcount\numsub
\numsec=0\numapp=0\numfig=1
\def\veroparagrafo{\number\numsec}\def\veraformula{\number\numfor}
\def\veraappendice{\number\numapp}\def\verasub{\number\numsub}
\def\verafigura{\number\numfig}

\def\section(#1,#2){\advance\numsec by 1\numfor=1\numsub=1%
\SIA p,#1,{\veroparagrafo} %
\write15{\string\Fp (#1){\secc(#1)}}%
\write16{ sec. #1 ==> \secc(#1)  }%
\hbox to \hsize{\titolo\hfill \number\numsec. #2\hfill%
\expandafter{\alato(sec. #1)}}\*}

\def\appendix(#1,#2){\advance\numapp by 1\numfor=1\numsub=1%
\SIA p,#1,{A\veraappendice} %
\write15{\string\Fp (#1){\secc(#1)}}%
\write16{ app. #1 ==> \secc(#1)  }%
\hbox to \hsize{\titolo\hfill Appendix A\number\numapp. #2\hfill%
\expandafter{\alato(app. #1)}}\*}

\def\senondefinito#1{\expandafter\ifx\csname#1\endcsname\relax}

\def\SIA #1,#2,#3 {\senondefinito{#1#2}%
\expandafter\xdef\csname #1#2\endcsname{#3}\else
\write16{???? ma #1#2 e' gia' stato definito !!!!} \fi}

\def \Fe(#1)#2{\SIA fe,#1,#2 }
\def \Fp(#1)#2{\SIA fp,#1,#2 }
\def \Fg(#1)#2{\SIA fg,#1,#2 }

\def\etichetta(#1){(\veroparagrafo.\veraformula)%
\SIA e,#1,(\veroparagrafo.\veraformula) %
\global\advance\numfor by 1%
\write15{\string\Fe (#1){\equ(#1)}}%
\write16{ EQ #1 ==> \equ(#1)  }}

\def\etichettaa(#1){(A\veraappendice.\veraformula)%
\SIA e,#1,(A\veraappendice.\veraformula) %
\global\advance\numfor by 1%
\write15{\string\Fe (#1){\equ(#1)}}%
\write16{ EQ #1 ==> \equ(#1) }}

\def\getichetta(#1){Fig. \verafigura%
\SIA g,#1,{\verafigura} %
\global\advance\numfig by 1%
\write15{\string\Fg (#1){\graf(#1)}}%
\write16{ Fig. #1 ==> \graf(#1) }}

\def\etichettap(#1){\veroparagrafo.\verasub%
\SIA p,#1,{\veroparagrafo.\verasub} %
\global\advance\numsub by 1%
\write15{\string\Fp (#1){\secc(#1)}}%
\write16{ par #1 ==> \secc(#1)  }}

\def\etichettapa(#1){A\veraappendice.\verasub%
\SIA p,#1,{A\veraappendice.\verasub} %
\global\advance\numsub by 1%
\write15{\string\Fp (#1){\secc(#1)}}%
\write16{ par #1 ==> \secc(#1)  }}

\def\Eq(#1){\eqno{\etichetta(#1)\alato(#1)}}
\def\eq(#1){\etichetta(#1)\alato(#1)}
\def\Eqa(#1){\eqno{\etichettaa(#1)\alato(#1)}}
\def\eqa(#1){\etichettaa(#1)\alato(#1)}
\def\eqg(#1){\getichetta(#1)\alato(fig. #1)}
\def\sub(#1){\0\palato(p. #1){\bf \etichettap(#1)\hskip.3truecm}}
\def\asub(#1){\0\palato(p. #1){\bf \etichettapa(#1)\hskip.3truecm}}

\def\equv(#1){\senondefinito{fe#1}$\clubsuit$#1%
\write16{eq. #1 non e' (ancora) definita}%
\else\csname fe#1\endcsname\fi}
\def\grafv(#1){\senondefinito{fg#1}$\clubsuit$#1%
\write16{fig. #1 non e' (ancora) definito}%
\else\csname fg#1\endcsname\fi}
\def\secv(#1){\senondefinito{fp#1}$\clubsuit$#1%
\write16{par. #1 non e' (ancora) definito}%
\else\csname fp#1\endcsname\fi}

\def\equ(#1){\senondefinito{e#1}\equv(#1)\else\csname e#1\endcsname\fi}
\def\graf(#1){\senondefinito{g#1}\grafv(#1)\else\csname g#1\endcsname\fi}
\def\secc(#1){\senondefinito{p#1}\secv(#1)\else\csname p#1\endcsname\fi}
\def\sec(#1){{\S\secc(#1)}}

\def\BOZZA{
\def\alato(##1){\rlap{\kern-\hsize\kern-1.2truecm{$\scriptstyle##1$}}}
\def\palato(##1){\rlap{\kern-1.2truecm{$\scriptstyle##1$}}}
}

\def\alato(#1){}
\def\galato(#1){}
\def\palato(#1){}

{\count255=\time\divide\count255 by 60 \xdef\hourmin{\number\count255}
        \multiply\count255 by-60\advance\count255 by\time
   \xdef\hourmin{\hourmin:\ifnum\count255<10 0\fi\the\count255}}

\def\oramin{\hourmin }

\def\data{\number\day/\ifcase\month\or gennaio \or febbraio \or marzo \or
aprile \or maggio \or giugno \or luglio \or agosto \or settembre
\or ottobre \or novembre \or dicembre \fi/\number\year;\ \oramin}
\setbox200\hbox{$\scriptscriptstyle \data $}
\footline={\rlap{\hbox{\copy200}}\tenrm\hss \number\pageno\hss}

 \let\b=\beta  \let\g=\gamma     \let\d=\delta  \let\e=\varepsilon
      \let\k=\kappa   \let\l=\lambda
\let\m=\mu                \let\p=\pi      
\let\s=\sigma \let\t=\tau        \let\c=\chi
   \let\o=\omega 
 \let\D=\Delta     \let\L=\Lambda

\def\\{\hfill\break} \let\==\equiv

\let\io=\infty 

\let\0=\noindent

\def\ie{\hbox{\it i.e.\ }}
\let\dpr=\partial 
\let\bs=\backslash

\def\tende#1{\,\vtop{\ialign{##\crcr\rightarrowfill\crcr
 \noalign{\kern-1pt\nointerlineskip}
 \hskip3.pt${\scriptstyle #1}$\hskip3.pt\crcr}}\,}
\def\otto{\,{\kern-1.truept\leftarrow\kern-5.truept\to\kern-1.truept}\,}
\def\fra#1#2{{#1\over#2}}

\def\PP{{\cal P}}\def\EE{{\cal E}}\def\VV{{\cal V}}
\def\FF{{\cal F}}\def\HH{{\cal H}}
\def\TT{{\cal T}}\def\NN{{\cal N}}\def\BB{{\cal B}}

\def\DD{{\cal D}}

\def\T#1{{#1_{\kern-3pt\lower7pt\hbox{$\widetilde{}$}}\kern3pt}}
\def\VVV#1{{\underline #1}_{\kern-3pt
\lower7pt\hbox{$\widetilde{}$}}\kern3pt\,}
\def\W#1{#1_{\kern-3pt\lower7.5pt\hbox{$\widetilde{}$}}\kern2pt\,}

\def\indica{\leaders \hbox to 0.5cm{\hss.\hss}\hfill}
\def\guida{\leaders\hbox to 1em{\hss.\hss}\hfill}
\mathchardef\oo= "0521

\def\xx{{\bf x}}
\def\yy{{\bf y}}\def\kk{{\bf k}}
\def\zz{{\bf z}}\def\uu{{\bf u}}
 \def\bP{{\bf P}}
\def\tt{{\bf t}}
\def\oo{{\underline \omega}}

\def\qed{\raise1pt\hbox{\vrule height5pt width5pt depth0pt}}

\def\indic{\hbox{\raise-2pt \hbox{\indbf 1}}}

\def\RRR{\hbox{\msytw R}}

\def\zzzz{\hbox{\msytww Z}}

%
%
%
\def\ins#1#2#3{\vbox to0pt{\kern-#2 \hbox{\kern#1 #3}\vss}\nointerlineskip}
%
%
%
\newdimen\xshift \newdimen\xwidth \newdimen\yshift

\def\insertplot#1#2#3#4#5{\par%
\xwidth=#1 \xshift=\hsize \advance\xshift by-\xwidth \divide\xshift by 2%
\yshift=#2 \divide\yshift by 2%
\line{\hskip\xshift \vbox to #2{\vfil%
#3 \includegraphics{#4.ps}}\hfill \raise\yshift\hbox{#5}}}

\def\initfig#1{%
\catcode`\%=12\catcode`\{=12\catcode`\}=12
\catcode`\<=1\catcode`\>=2
\openout13=#1.ps}

\def\endfig{%
\closeout13
\catcode`\%=14\catcode`\{=1
\catcode`\}=2\catcode`\<=12\catcode`\>=12}


\initfig{fig51}
\write13<
\write13<
\write13<gsave .5 setlinewidth 40 20 260 {dup 0 moveto 140 lineto} for stroke
grestore>
\write13</punto { gsave  
\write13<2 0 360 newpath arc fill stroke grestore} def>
\write13<40 75 punto>
\write13<60 75 punto>
\write13<80 75 punto>
\write13<100 75 punto 120 68 punto 140 61 punto 160 54 punto 180 47 punto 200
40 punto>
\write13<220 33 punto 240 26 punto 260 19 punto>
\write13<120 82.5 punto>
\write13<140 90 punto>
\write13<160 80 punto>
\write13<160 100 punto>
\write13<180 110 punto>
\write13<180 70 punto>
\write13<200 60 punto>
\write13<200 120 punto>
\write13<220 110 punto>
\write13<220 50 punto>
\write13<240 100 punto>
\write13<240 60 punto>
\write13<120 50 punto>
\write13<260 20 punto>
\write13<240 40 punto>
\write13<240 50 punto>
\write13<260 70 punto>
\write13<200 80 punto>
\write13<260 90 punto>
\write13<260 110 punto>
\write13<220 130 punto>
\write13<40 75 moveto 100 75 lineto 140 90 lineto 200 120 lineto 220 130
lineto>
\write13<200 120 moveto 240 100 lineto 260 110 lineto>
\write13<240 100 moveto 260 90 lineto>
\write13<140 90 moveto 180 70 lineto 200 80 lineto>
\write13<180 70 moveto 220 50 lineto 260 70 lineto>
\write13<220 50 moveto 240 40 lineto>
\write13<220 50 moveto 240 50 lineto>
\write13<100 75 moveto 260 20 lineto>
\write13<100 75 moveto 120 50 lineto stroke>
\write13<grestore>
\endfig
\openin14=\jobname.aux \ifeof14 \relax \else
\input \jobname.aux \closein14 \fi
\openout15=\jobname.aux
\footline={\tenrm\hss \number\pageno\hss}
\def\*{\vskip0.3cm}

{\baselineskip=12pt
\centerline{\titolone Gap generation in the BCS model} 
\centerline{\titolone with finite range temporal
interaction}
\vskip1.truecm
\centerline{\titolo{Vieri Mastropietro}}
\vskip.5cm
\centerline{Dipartimento di Matematica, Universit\`a di Roma ``Tor
Vergata''}
\centerline{Via della Ricerca Scientifica, I-00133, Roma}
\vskip1cm \0{\cs Abstract.} {\it 
In the [BCS] paper the theory
of superconductivity was developed for the BCS model,
in which the (instantaneous) 
interaction is only between fermions of opposite momentum and spin. 
Such model was analyzed by variational methods, 
finding that a superconducting behavior is energetically 
favorable. Subsequently it was claimed that
in the thermodynamic limit the BCS model
is equivalent to the (exactly solvable) quadratic mean field BCS model;
a rigorous proof of this claim is however still lacking.
In this paper we consider the  
BCS model with a finite range temporal
interaction, and we
prove rigorously its equivalence with the mean field BCS model
in the thermodinamic limit
if the range is long enough,
by a (uniformly convergent) perturbation expansion about mean field theory.}
\vskip1cm 
\vskip.5cm
\section(1,Introduction and main results)
\vskip.5cm 

Bardeen, Cooper and Schreifer [BCS] developed their theory 
describing superconductors by 
the {\it BCS model}, in which
the interaction has infinite range and 
only fermions of opposite momentum and spin ({\it Cooper pairs})
interact; the Hamiltonian of this model is
$$H_{BCS}=\sum_\s\int_V  d{\vec x} a^+_{{\vec x},\s} 
\left(-{\partial^2_{\vec x}\over 2 m}\right)
a^+_{{\vec x},\s}-{\l\over V}[\sum_\s \int_V 
d{\vec x} a^+_{{\vec x},\s}
a^+_{{\vec x},-\s}][\sum_\s\int_V d{\vec y} a^-_{{\vec y},\s}
a^-_{{\vec y},-\s'}]\Eq(ggh)$$
where $a^\pm_{x,\s}$ are creation
or annihilation fermionic field operators with spin $\s$
in a $d$-dimensional box with side $L$ and $V=L^d$, $m$
is the mass and $\l>0$ is the (attractive) coupling.
By a variational procedure it was found that it was
energetically favorable to form a superconducting phase.
Later on it was realized that 
the properties of such superconducting phase are identical to the ones of 
the {\it mean field BCS} model, an exactly solvable model
in which 
the interaction is quadratic and the Hamiltonian has the form
$$H_{MF}=|\D|^2+\sum_\s\int_V  d{\vec x} a^+_{{\vec x},\s} 
\left(-{\partial_{\vec x}^2\over 2 m}\right)
a^+_{{\vec x},\s}-
\sqrt{\l} \D[\sum_\s
\int_V d{\vec y} a^-_{{\vec y},\s}
a^-_{{\vec y},-\s}]-\sqrt{\l} \bar\D[\sum_\s \int_V d{\vec y} a^+_{{\vec y},\s}
a^-_{{\vec y},-\s}]\Eq(gg4)$$ 
where $\D$ is a complex number to be
determined minimizing the ground
state energy (that is $\D$ solves the BCS gap equation). 
It has been argued in several papers, 
starting from [BR],[B],[H],that 
in the limit $V\to\io$ the reduced BCS model 
\equ(ggh) and the mean field model \equ(gg4) 
{\it have the same correlation functions}; this seems quite
natural also by analogy with lattice classical
statistical mechanics in which infinite range interaction gives mean field behavior
in the thermodynamic limit.
Indeed many arguments has been given to support this claim
in the last fifty years but, as far as I known, a rigorous proof is still
lacking; aim of this paper is show that a simple proof of this
claim can be given at least 
if the instantaneous interaction in the reduced
BCS model \equ(ggh) is replaced with a long (but finite) range time interaction

We consider then a generalization of the reduced BCS model
in which fermions are on 
on a cubic lattice with step 1
and a {\it time-dependent} interaction between Cooper pairs
is considered; indeed, as stressed for instance in [CEKO],
a realistic model for superconductivity
should include a bosonic Hamiltonian describing phonons
and a boson-fermion interaction, which
can be written in a purely fermionic model 
only if {\it time dependent} interaction between fermions is included.
The two point Schwinger
function of the reduced BCS model on a lattice
with time dependent interaction 
can be written as
{\it Grassmann functional integral} in the following way
$$<\psi^\e_{\kk,\s}\psi^{\e'}_{\kk',\s'}>_{L,\b,M}={\int P(d\psi) e^{-\VV-
h\sum_{\e,\s} 
\int d\xx\psi^\e_{\xx,\s}\psi^\e_{\xx,-\s}}
\psi^\e_{\kk,\s}\psi^{\e'}_{\kk',\s}\over\int P(d\psi)
e^{-\VV-h\sum_{\s,\e} \int d\xx
\psi^\e_{\xx,\s}\psi^\e_{\xx,-\s}
}}
\Eq(h1)$$
where $\int d\xx=\int_{-{\b\over 2}}^{\b\over 2} dx_0\sum_{x\in\L}$
and $\L$ is a $d$-dimensional lattice with  step $1$ and
$$\VV={\l\over L^d}\sum_{\s,\s'}\int
d\xx\int d\yy v(x_0-y_0) \psi^+_{\xx,\s}
\psi^+_{\xx,-\s}\psi^-_{\yy,-\s}\psi^-_{\yy,-\s'}\Eq(2ss)$$
In the above expression $\{\psi^\pm_{\kk,\s}\}$
is a set of {\it Grassmannian variables}, $\kk\in {\cal D}_{L,\b}$
where
${\cal D}_{L,\b}\={\cal D}_L
\times {\cal D}_\b$, with ${\cal D}_L\=\{k={2\pi n/L}, n\in \zzzz^d, 
-[L/2]\le n_i \le [(L-1)/2]\}$ and
${\cal D}_\b\=\{k_0=2(n+1/2)\pi/\b, n\in
Z, -M\le n \le M-1\}$, and
$P(d\psi)$ is a linear functional on the
generated Grassmann algebra  such that
$$\int P(d\psi) \hat \psi^-_{\kk_1,\s_1}
\hat \psi^+_{\kk_2,\s_2} = L^d\b \d_{\kk_1,\kk_2}\d_{\s_1,\s_2}
\hat g(\kk_1)\;,\quad \hat g(\kk)= {1\over -ik_0+\e(k)-\m}
\; .\Eq(2.8)$$
where 
$$\e(\vec k)=\sum_{i=1}^d(1-\cos k_i)\Eq(2a)$$
is the {\it dispersion relation} and $\m$ is the chemical potential.
We define also {\sl Grassmannian field} $\psi^\pm_\xx$ is defined by, 
if $\xx=(x,x_0)$ with $x_0\in (-{\b\over 2},{\b\over 2}]$
and $x=(x_1,..,x_d)$ with $x_i=1,2,..,L$, 
$$\psi_{\xx,\s}^{\pm}= {1\over L^d\b} \sum_{\kk\in {\cal D}_{L,\b}}\hat \psi_{\kk,\s}^{\pm}
e^{\pm i\kk\cdot\xx}\Eq(chic)$$
The external field $h$ is introduced to break the number
symmetry and which will be removed 
after the thermodynamic limit $L\to\io$ will be taken
and  $v(x_0-y_0)$ is a {\it Kac potential} 
with a {\it long but finite} range potential $\k^{-1}$; for definiteness
we choose
$$v(t)={1\over\b}\sum_{k_0={2\pi n_0\over\b}\atop n_0=0,\pm 1,\pm 2,..,\pm M} 
e^{-i k_0 t}{\k^2\over k_0^2+\k^2}\Eq(p2)$$
Finally $M$ is an ultraviolet cutoff in the time direction introduced 
to make the Grassmann integral well defined, and
and {\it the limit $M\to\io$ must be taken before the thermodynamic limit
$V\to\io$}.

As we mentioned above, it was claimed in [BR] that
the reduced BCS model (not solvable) 
should be equivalent (in the sense that the Schwinger functions 
coincide) to the mean field BCS model (solvable) in the limit
$L\to\io$. In [BZT] indeed it was shown at each order
of the perturbative expansion that the difference of 
the correlation functions between the reduced and the BCS model
goes as $O(V^{-1})$ at each order, but to make this argument rigorous
one has to prove the uniformity of 
the convergence of the perturbative expansion. A similar
perturbative argument in a more modern (RG) language 
has been given in [SHML], in which it is pointed out the similarity
of the perturbative expansion of the reduced BCS model with the 
so called $O({1\over N})$ expansion. In 
[B] and [H] the proof of such equivalence
was based on the idea 
that the spatial averages of field operators like $V^{-1}\int d{\vec x} 
a^+_{\s,\vec x}a^+_{-\s,\vec x}$ 
may be substituted by numbers in the thermodynamic limit, 
since commutators with them has an extra one volume factor. 
Such a proof was criticized by several authors;
for instance in [TW] it was shown that
the convergence of the reduced BCS to the mean field
model is true only in a 
rather small subspace and not in general.
In [M] a new proof of the equivalence based on a functional integral
approach was given, but in the analysis involves unjustified exchange
of the $L\to\io$ limit with the $M\to\io$ limit. 
Finally
in [T] a correct proof of such equivalence was given, 
but only under that rather unrealistic assumption the
{\it the dispersion
relation is a constant} (degenerate BCS model). 

It is apparently surprising the difficulty in
proving that a infinite range interaction interaction like the one 
in \equ(ggh) leads to a mean field behavior in the thermodinamic limit; 
indeed in classical
statistical mechanics for spin lattice systems the proof
of a similar statement
is a two line computation. The difficulty in the quantum case
can be clearly understood in 
the functional integral formulation \equ(h1);
in such a representation {\it 
the interaction $\VV$ is not factorized} 
contrary to what happen in the Hamiltonian formulation,
and this make the model {\it not} exactly solvable.
Of course by replacing $v(x_0-y_0)$ in \equ(h1)
with a constant (that is we consider
an {\it infinite} 
long range time interaction $\k^{-1}=\io$) 
the interaction in the functional integral is factorized 
and the model is exactly solvable; mean field behavior in the thermodinamic limit is then easily established,
by performing a saddle point analysis essentially identical
to the one for long range spin systems, see [L].

Aim of this paper is to prove that even if the range $\k^{-1}$ in \equ(p2) 
is {\it finite},
so that the interaction is not factorized and the model {\it not} solvable,
the BCS model \equ(h1) is equivalent to the mean field BCS model
if $\k$ is large enough, in the limit $V\to\io$; that is the BCS model has a 
phase transition into a superconducting state described by the BCS theory.

Our main result is the following.
\vskip.5cm
{\bf Theorem} {\it Assume $\m<2$ and $\l>0$; 
there exist $\b_c(\l)$ and
$\k_0(\b)>0$ such that
for $\b\ge\b_c(\l)$ and $0<\k<\k_0(\b)$ the Schwinger functions
\equ(h1) with interaction \equ(p2)
are such that
$$\lim_{h\to 0^+}
\lim_{L\to\io}<\psi^-_{\kk,\s}\psi^+_{\kk,\s}>
={-i k_0+\e(\vec k)-\m
\over k_0^2+(\e(\vec k)-\m)^2+\l|\D|^2}\Eq(ppo1)$$
$$\lim_{h\to 0^+}
\lim_{L\to\io}<\psi^+_{\kk,\s}\psi^+_{-\kk,\s}>=
{\sqrt{\l}\D 
\over k_0^2+(\e(\vec k)-\m)^2+\l |\D|^2}
\Eq(ppll1)$$
where $\D\equiv\D(\b)$ is the positive solution 
of the BCS gap equation
$$1=\l\int {d{\vec k}\over (2\pi)^d} {tangh({\b\over 2}\sqrt{E^2({\vec k})+
\l\D^2})\over 2((\e({\vec k})-\m)^2+\l\D^2)}\Eq(BCS)$$
and $\b_c(\l)$ is the minimal $\b$ for which
\equ(BCS) admits a solution.
}
\vskip.5cm
The above Theorem ensures that, at a fixed temperature $\b$
and for range $\k^{-1}$ large enough,
the BCS model has the same behavior of the BCS mean field model;
in particular for $\b\ge \b_c$ a gap is generated and the particle number
symmetry is broken as $<\psi^+_{\kk,\s}\psi^+_{-\kk,\s}>$
is different from zero; this means that there is a phase transition into
a superconducting phase for temperatures low enough.
As an immediate corollary, it follows that for an interaction like 
\equ(p2) with an exponentially large range $O(e^{{a\over \l}})$
a gap is generated
and the particle number symmetry is broken for temperatures  small enough.

The proof of the above statement is by perturbation theory
about the mean field theory,
using as a perturbative parameter the inverse range $\k$ of the Kac potential 
\equ(p2); this is a
a classical approach
in classical statistical mechanics to prove phase transition 
beyond mean field theory, see for instance [LMP].
We will show that the correction with respect to the mean
field Schwinger function is expressed by a {\it convergent}
series expansion (uniformly in the volume)
and each order is $O(V^{-1})$;
uniform convergence is established via determinant bounds for fermions.
We can prove convergence only for small $\k$, as it turns out that
$\k_0=O(\b^{-{d+6\over 2}})$; of course 
it would be very interesting to prove convergence
up to $\k_0=O(1)$ or even for any $\k$, so otaining a
real solution of the BCS model with instantaneous interaction.
\vskip.5cm
\section(2,Partial Hubbard-Stratonovich transformation)
\vskip.5cm
%
%
In momentum space we can write the interaction $\VV$ 
in the following way
$$\VV= -{\l\over (\b V)^3}\sum_{\kk,\kk'}
\sum_{p_0={2\pi\over\b}(n_0+1)\atop n_0\in Z} 
\sum_{\s,\s'}v(p_0)\psi^+_{\s,\kk}\psi^+_{-\s,-\kk+p_0}
\psi^-_{\s',\kk'}\psi^-_{-\s',-\kk'+p_0}\Eq(intt)$$
where we have used that $p_0=k_{0,1}+k_{0,2}={2\pi\over\b}
((n_{0,1}+n_{0,2})+1)$.
We split the interaction $\VV$ as sum over two terms 
$$\VV=\bar\VV+\hat\VV_2\Eq(p11)$$
$$\bar\VV=-{\l\over (\b V)^3}\sum_{\kk,\kk'}\sum_{\s,\s'}\psi^+_{\s,\kk}\psi^+_{-\s,-\kk}
\psi^-_{\s',\kk'}\psi^-_{-\s',-\kk'}\Eq(p12)$$

$$\hat\VV= -{\l\over (\b V)^3}
\sum_{\kk,\kk',|p_0|\ge {2\pi\over\b}}\sum_{\s,\s'} 
v(p_0)\psi^+_{\s,\kk}\psi^+_{-\s,-\kk+p_0}
\psi^-_{\s',\kk'}\psi^-_{-\s',-\kk'+p_0}\Eq(p13)$$
%
%
Note that $\bar\VV$ can be written as, $\e=\pm$
$$\bar\VV=-\D^+\D^-\quad\quad\D^\e\equiv 
{\sqrt{\l}\over (\b V)^{1/2}}\DD^\e=
{\sqrt{\l} \over (\b V)^{3/2}}
\sum_{\s}\sum_{\kk}\psi^\e_{\kk,\s}
\psi^{\e}_{-\kk,\s}\Eq(p80)$$
that is can be written as the product of the total number
of Cooper pairs. 
Let us consider the {\it generating function}
of the Schwinger functions
$$e^{{\cal S}_{L,\b,h}(J)}=
\int P(d\psi)e^{2 
\D^+\D^--\hat\VV}
e^{-h{\sqrt{\b V}\over\sqrt{\l}}
\D^+-h{\sqrt{\b V}
\over\sqrt{\l}}\D^-}e^{\int d\xx\sum_\s[J^+_{\xx,\s}
\psi^-_{\xx,\s}+\psi^+_{\xx,\s}
J^-_{\xx,\s}]}
\Eq(bo2)$$
where $J^\pm$ are external Grasssmann field, so that 
$$<\psi^{\e}_{\xx,\s}\psi^{\e'}_{\yy,\s}>
={\partial^2\over\partial J^\e_\xx \partial J^{\e'}_\yy}
S(J)|_{J=0}\Eq(nnmm)$$
By using the identity ({\it Hubbard-Stratanovich
transformation}) ($\phi=u+iv$, $\bar\phi=u-iv$, $u,v\in R$)
$$
e^{2ab}={1\over 2\pi}\int_{R^2} du dv e^{-{1\over 2}|\phi|^2}
e^{a\phi+b\bar\phi}\Eq(z1)$$
we can rewrite the above expression as
$$e^{{\cal S}_{L,\b,h}(J)}=
{1\over 2\pi}\int_{R^2} du dv 
e^{-{1\over 2}|\phi|^2}
\int P(d\psi)e^{-\hat \VV}
e^{(\phi-h{\sqrt{\b V}\over\sqrt{\l}})\D^+ +(\bar\phi-h{\sqrt{\b V}
\over\sqrt{\l}})\D^-}
e^{\int d\xx\sum_\s[J^+_{\xx,\s}
\psi^-_{\xx,\s}+\psi^+_{\xx,\s}
J^-_{\xx,\s}]}
\Eq(z2)$$
Performing the change
of variables
$ (u,v)\to \sqrt{\b V} (u,v)$
we obtain
$$e^{{\cal S}_{L,\b,h}(J)}=
{\b V\over 2\pi}\int_{R^2} du dv 
e^{-{\b V\over 2}(v^2+(u+{h\over \sqrt{\l}})^2}
e^{-\b V\FF_{L,\b,h}(u,v)+\BB_{L,\b,h}(u,v,\phi)}\Eq(kkll)$$
where
$$e^{-\b V\FF_{L,\b,h}(u,v)+\BB_{L,\b,h}(u,v,J)}=
\int P(d\psi)
e^{-\hat\VV}
e^{\sqrt{\l}\phi\DD^+
+\sqrt{\l}\bar\phi\DD^-}
e^{\int d\xx\sum_\s[J^+_{\xx,\s}
\psi^-_{\xx,\s}+\psi^+_{\xx,\s}
J^-_{\xx,\s}]}
\Eq(z3)$$
%
%
%
and by definition $\BB_{L,\b,h}(u,v,J)$ is vanishing for $J=0$
so that $\FF_{L,\b,h}(u,v)$ is given by
$$e^{-\b V\FF_{L,\b,h}(u,v)}=
\int P(d\psi)
e^{-\hat\VV}
e^{\sqrt{\l}\phi\DD^+
+\sqrt{\l}\bar\phi\DD^-}
\Eq(zii)$$
We are interested in computing the two point Scwhinger
function, given by \equ(nnmm)
$$<\psi^{\e}_{\kk,\s}\psi^{\e'}_{-\e\e'\kk,\s}>={1\over Z_{L,\b,h}}
\int_{R^2} du dv 
e^{-{\b V\over 2}(v^2+(u+{h\over \sqrt{\l}})^2)}
e^{-\b V\FF_{L,\b,h}(u,v)} S_{L,\b}^{\e,\e'}(\kk,u,v)\Eq(kkll)$$
where $S_{L,\b,h}(u,v)=\partial_{J_\xx^\e}\partial_{J_\yy^{\e'}}
\BB(J,u,v)|_{J=0}$ and 
$$Z_{L,\b,h}=\int_{R^2} du dv 
e^{-{\b V\over 2}(v^2+(u+{h\over \sqrt{\l}})^2}
e^{-\b V\FF_{L,\b,h}(u,v)}\Eq(nn)$$
We will show in the following section that 
$$\FF_{\L,\b,h}(u,v)=t_{BCS}+\bar\FF_{L,\b,h}(u,v)\Eq(fff)$$
where, if $E({\vec k})=\e(\vec k)-\m$
$$t_{BCS}={1\over V}\sum_{\vec k} 2\b^{-1}
\log {cosh({\b\over 2}\sqrt{E^2({\vec k})+\l|\phi|^2})\over
cos h {\b\over 2} E({\vec k})}\Eq(bbn)$$
is the free energy in the mean field BCS model [BCS] and
$\bar\FF_{L,\b,h}$ is the perturbation to the mean field;
we will show in the following section that, for $0<\k<\k_0(\b)$,
$\k_0(\b)=C^{-1}\b^{-d-6\over 2}$, for a suitable constant $C$
$$|\bar\FF_{L,\b,h}(u,v)|\le C {\l\over V}(\k^2\b^3) \b^{d+2}\Eq(ggfg)$$
hence $V\bar\FF_{L,\b,h}(u,v)$ it is uniformly bounded as $V\to\io$;
it is more convenient to call $V\bar\FF_{L,\b,h}(u,v)\equiv 
\hat\FF_{L,\b,h}(u,v)$ and 
we can write the two point Schwinger functions as
$${1\over Z_{L,\b,h}}
\int_{R^2} du dv 
e^{-{\b V\over 2}[v^2+(u+{h\over \sqrt{\l}})^2+t_{BCS}(u,v)]}
e^{-\b\hat\FF_{L,\b,h}(u,v)}
S_{L,\b}^{\e,\e'}(\kk,u,v)\Eq(kkll)$$
By the saddle point Theorem, for $\b$ large enough
$$\lim_{L\to\io} {e^{-\b L (v^2+(u+{h\over \sqrt{\l}})^2+t_{BCS}(u,v))}\over 
\int du dv e^{-\b L (v^2+(u+{h\over \sqrt{\l}})^2+t_{BCS}(u,v))}}=\d(u)\d(v-v_0)\Eq(dis)$$
where $v_0$ is given by the negative (for $h>0$) solution of
$$v_0[\l\int {d{\vec k}\over (2\pi)^d}
{tangh({\b\over 2}\sqrt{E^2(k)+
\l v_0^2})\over 2(E^2({\vec k})+\l v_0^2)}]=2 |h|
\Eq(bcs22)$$
In the limit $h\to 0$ it reduces to the BCS equation \equ(BCS).
Moreover we will show in the following section that 
$S_{L,\b}^{\e,\e'}-S_{L,\b}^{\e,\e',BCS}$ is $O(V^{-1})$
so that the Theorem follows.
\*
\vskip.5cm
\section(16,Convergence of series expansion)
\vskip.5cm
\sub(1.1akl){\it The partition function}

We can ``absorb'' the quadratic fermion term in the the free interaction
$$\int P(d\psi)e^{\sqrt{\l}\phi\DD^+
+\sqrt{\l}\bar\phi\DD^-}e^{-\hat\VV(\psi)}=
e^{-\b V t_{BCS}}
\int P_{\s}(d\psi)e^{-\hat\VV(\psi)}\Eq(fs)$$
where 
$$\hat\VV(\psi)= -{\l\over V}\sum_{\s,\s'}
\int d\xx d\yy \tilde v(x_0-y_0) \psi^+_{\s,\xx}\psi^+_{-\s,\xx}
\psi^-_{\s',\yy}\psi^-_{-\s',\yy}\Eq(p14)$$
and
$$\tilde v(x_0-y_0)={1\over \b}\sum_{k_0\not=0} e^{i k_0 t} 
{\k^2\over \k^2+k_0^2}\Eq(int)$$
and, if $\s=\sqrt{\l}\phi$
$$P_\s(d\psi) =
\prod_{\kk}
{d\hat\psi^{+}_{\kk}d\hat\psi^{-}_{\kk'}\over \NN(\kk)}
\left\{-{1\over V\b} \sum_{\kk'}
\sum_{\e,\e'=\pm} \hat\psi^{\e }_{\e\kk} T_{\e,\e'}
\hat\psi^{\e'}_{\e'\kk}\right\}\Eq(2.66)$$
where $\NN(\kk)$ is the normalization of $P_\s(d\psi)$
and 
$$t_{BCS}=-{1\over V\b}\sum_{\kk}\log {k_0^2+E^2(k)+|\s|^2
\over k_0^2+E^2(k)}\Eq(jhhg)$$
and the $2\times2$ matrix $T(\kk')$ is given by
$$T(\kk) = \left(\matrix{-ik_0+ E(k) & \sigma \cr
\bar\sigma& -ik_0-E(k) \cr}\right)\;.\Eq(2.69)$$
We can equivalently write, see for instance [L], $t_{BCS}$ as \equ(bbn)
and of course $t_1\le \sqrt{|\l|} C [1+|\phi|]$. 
The propagator of $P_\s(d\psi)$
is given by
$$\int P_\s(d\psi)\psi^\e_{\xx,\s}\psi^{\e'}_{\yy,\s}\equiv
g_{\e,\e'}(\xx,\yy)
={1\over V\b}\sum_\kk e^{-i\kk(\xx-\yy)}[T^{-1}(\kk)]_{\e,\e'}\Eq(prp)$$
We decompose the free
propagator $\hat g_\kk$ into a sum of two propagators supported in
the regions of $k_0$ ``large'' and ``small'', respectively. The
regions of $k_0$ large and small are defined in terms of a smooth
support function $H_0(t)$, $t\in\RRR$, such that   
$$H_0(t) = \cases{
1 & if $t <1/\g \;,$ \cr 0 & if $t >1\;,$\cr}\Eq(2.9)$$
with $\g>1$. We define  
$h(k_0)= H_0(|k_0|)$ 
so that we can rewrite $\hat g_\kk$ as:
$$\hat g_\kk=\hat g^{(u.v.)}(\kk)+\hat g^{(i.r.)}(\kk)\Eq(2.10)$$
where
$$g^{(i.r.)}_{\e,\e'}(\xx,\yy)
={1\over V\b}\sum_\kk e^{-i\kk(\xx-\yy)} h(k_0) [T^{-1}(\kk)]_{\e,\e'}\Eq(prp1)$$
$$g^{(u.v.)}_{\e,\e'}(\xx,\yy)={1\over V\b}\sum_\kk 
e^{-i\kk(\xx-\yy)} (1-h(k_0)) 
[T^{-1}(\kk)]_{\e,\e'}\Eq(prp2)$$
In the Appendix we show that 
$$\int P_{\s}(d\psi^{u.v.})e^{-\hat\VV(\psi^{i.r.}
+\psi^{u.v})}=e^{-\VV^0(\psi^{i.r.})}
\Eq(fs1)$$
with
$$\VV^0=-{\l\over V}\sum_{\s,\s'}
\int d\xx d\yy \tilde v(x_0-y_0)
\psi^+_{\xx,\s}\psi^+_{\xx,-\s}
\psi^-_{\yy,\s'}\psi^-_{\yy,-\s'}+
\sum_{n=1}^\io \int d\xx_1...\int d\xx_{2n} 
W^0_{2 n}(\xx_1,..,\xx_{2n})\prod_{i=1}^{2n} 
\psi^{\e_i}_{\xx_i,\s_i}\Eq(ll1)$$
with
$$\fra{1}{V\b}\int d\xx_1\cdots d\xx_{2n}
|W^{(0)}_{n}(\xx_1,..,\xx_{2n})|
\le C^n |\l|^{max(1,n-1)}(\k^2\b^3)^{max(1,n-1)}
\Eq(111.6)$$ 
\vskip.5cm
\sub(1.1akl1){\it Convergence of the infrared integration}

We define a distance ${\bf d(\xx,\yy)}_{L,\b}=(d_\b(x_0,y_0),
d_L(x_1,y_1),...,d_L(x_n,y_n)$ as
$$d_\b(x_0,y_0)={\b\over\pi}\sin {\pi\over\b}(x_0-y_0)
\quad d_L(x_i,y_i)={L\over\pi}\sin {\pi\over L}(x_i-y_i)\Eq(lkn)$$
In order to perform the infrared integration we 
need the large distances behaviour of the infrared propagator.
\vskip.5cm
{\bf Lemma} {\it For any integer $N$ the following bounds hold
$$|g_{\e,-\e}^{(i.r.)}(\xx,\yy)|\le \b {C_N\over 1+
[\b^{-1} {\bf d}(\xx-\yy)]^N}\Eq(pl11)$$
$$|g_{\e,\e}^{(i.r.)}(\xx,\yy)|\le \b {\sqrt{\l}|\phi|\over
\sqrt{\l}|\phi|+\b^{-1}}
{C_N\over 1+
[\b^{-1} {\bf d}(\xx-\yy)]^N}\Eq(pl11)$$}
\vskip.5cm
{\it Proof.} The above bounds follows by integrating by parts.
Consider integers $N_0,N_1,..,N_d$
note that, $i=1,..,d$
$$\eqalign{
&d_L(x_i,y_i)^{N_i} d_\b(x_0,y_0)^{N_0} 
g^{i.r}_{\e,\e'}(\xx-\yy) =\cr
&e^{-i\p (xL^{-1}N_i+x_0\b^{-1}N_0)}(-i)^{N_0+N_i}
{1\over V\b}\sum_{\kk}e^{-i\kk(\xx-\yy)} 
\dpr_{k}^{N_i} \dpr_{k_0}^{N_0}
\left[(1-h(k_0))[T_{0}^{-1}(\kk')]_{\e,\e'}\right]\;,\cr}\Eq(2.105)$$
where $\dpr_{k}$ and $\dpr_{k_0}$ denote the discrete derivatives.
The bound then easily follows noting that 
$T_{0}^{-1}(\kk')]_{\e,\e'}$ is bounded by $C\b$ and 
each derivatives over it is bounded by an extra $\b$. The non diagonal
term has an extra ${|\s|\over \kk^2+|\s|^2}$
in the bound, from which we see that the bound is 
uniform in $\s$.).\qed 
\vskip.5cm
We can write
$$\int P(d\psi^{(i.r.)}) e^{-\VV^0(\psi^{i.r.})}=
e^{\sum_{n=0}^{\io}(-1)^n {1\over n!}\EE^T(\VV^0,...;\VV^0)}\Eq(ass)$$
where $\EE^T$ are the fermionic truncated expectations
$$\EE^T_\psi(\psi;n)={\partial^n\over\partial\l^n}\log\int P(d\psi)
e^{\l X(\psi)}\Big|_{\l=0}\Eq(7.19)$$
We write \equ(ll1) as
$$\sum_{P} \int d\xx_{P} W(\xx_{P})
\tilde\psi(P)\Eq(jjmb)$$
where $P_v$ is the set of field labels appearing
in \equ(ll1), $W(\xx_{P_v})$ are the kernels in 
\equ(ll1), that is $V^{-1}\tilde v(x_0-y_0)$
or $W(\xx_1,..,\xx_{2n})$ and
$$\tilde\psi(P)=\prod_{f\in P}\psi^{\e(f)}_{\xx(f),\s(f)}
\Eq(nhds)$$
Then we get
$$\EE^T(\VV^0;n)=\sum_{P_{1},...,P_{n}}
\int d\xx_{P_{1}}...\int d\xx_{P_{n}} 
W(\xx_{P_{1}})...W(\xx_{P_{n}})
\EE^T(\tilde\psi(P_{1})....\tilde\psi(P_{n}))
\Eq(tt)$$
The fermionic truncated 
expectations can be bounded by the formula
(see [GM] for example), if $s>1$,
$$\tilde\EE^T(\tilde\psi (P_1),...,\tilde\psi(P_s))=
\sum_{T}\prod_{l\in T} g_{\e_l,\e'_l}(\xx_l-\yy_l)\int
dP_{T}(\tt) \det G^{T}(\tt)\;,\Eq(3.74)$$
where 
$$\tilde\psi(P)=\prod_{f\in P}\psi_{\xx(f),\s(f)}^{\e(f)}\Eq(ddf)$$
and
\vskip.5cm
a) $T$ is a set of lines forming an {\it anchored tree} between
the cluster of poins $P_1,..,P_s$ \ie $T$ is a set
of lines which becomes a tree if one identifies all the points
in the same clusters.
\vskip.5cm
c) $\tt=\{t_{i,i'}\in [0,1], 1\le i,i' \le s\}$, $dP_{T}(\tt)$
is a probability measure with support on a set of $\tt$ such that
$t_{i,i'}=\uu_i\cdot\uu_{i'}$ for some family of vectors $\uu_i\in \RRR^s$ of
unit norm.
\vskip.5cm
d) $G^{T}(\tt)$ is a $(N-s+1)\times (N-s+1)$ matrix, $2N=|P_1|+..+|P_s|$
whose
elements are given by $G^{T}_{ij,i'j'}=t_{i,i'}
g_{\o^-,\o^+}(\xx_{ij}-\yy_{i'j'})$ with
$(f^-_{ij}, f^+_{i'j'})$ not belonging to $T$.
\vskip.5cm
If $s=1$ the sum over $T$ is empty, but we can still
use the above equation by interpreting the r.h.s.
as $1$ if $P_1$ is empty, and ${\rm det} G(P_1)$ otherwise.

We bound the determinant using the well known
{\it Gram-Hadamard inequality}, stating
that, if $M$ is a square matrix with elements $M_{ij}$ of the form
$M_{ij}=<A_i,B_j>$, where $A_i$, $B_j$ are vectors in a Hilbert space
with
scalar product $<\cdot,\cdot>$, then
$$|\det M|\le \prod_i ||A_i||\cdot ||B_i||\;.\Eq(7.27)$$
where $||\cdot||$ is the norm induced by the scalar product.

Let $\HH=\RRR^s\otimes \HH_0$, where $\HH_0$ is the Hilbert space of complex
four dimensional vectors $F(\kk)=(F_1(\kk),\ldots,F_4(\kk)$), $F_i(\kk)$
being a function on the set $\DD_{-,-}$, with scalar product
$$<F,G>=\sum_{i=1}^4 {1\over L\b}\sum_{\kk} F^*_i(\kk) G_i(\kk)\;.
\Eq(7.28)$$
and one checks that
$$G^{T}_{ij,i'j'}=t_{i,i'}
g^{(\chi)}_{\o^-_l,\o^+_l}(\xx_{ij}-\yy_{i'j'})
=<\uu_i\otimes A_{\xx(f^-_{ij}),\o(f^-_{ij})},
\uu_{i'}\otimes B_{\xx(f^+_{i'j'}),\o(f^+_{i'j'})}>\;,\Eq(7.29)$$
where $\uu_i\in \RRR^s$, $i=1,\ldots,s$, are the vectors such that
$t_{i,i'}=\uu_i\cdot\uu_{i'}$, and
$$\eqalign{
A_{\xx,\o}(\kk)&=e^{i\kk'\xx}{\sqrt{h(k_0)} \over
\sqrt{k_0^2+E^2+|\s|^2}} \cdot \cases{
(-i k_0+E(k),0,\s,0),& if $\o=+1$,\cr
(0,\bar\s,0,1),& if $\o=-1$,\cr}\cr
B_{\xx,\o}&=e^{i\kk'\yy}{\sqrt{h(k_0)}\over
\sqrt{k_0^2+E^2+|\s|^2}} \cdot \cases{
(1,1,0,0),& if $\o=+1$,\cr
(0,0,1,-i k_0-E(k))),& if $\o=-1$.\cr}\cr}\Eq(7.30)$$

Hence from \equ(7.27), as $||A||\le C$ and $||B||\le \b$ 
we find
$$|G^{T}_{ij,i'j'}|\le C_1^{N-s+1}\b^{N-s+1}\Eq(7.31)$$
where $C_1$ is an $O(1)$ constant.  

By using the above formula in \equ(tt) we get
$$|\EE^T(\VV^0,..,\VV^0)|\le \sum_{P_{1},...,P_{n}}\b^{{1\over 2}[|P_1|+..+|P_n|]}
\int d\xx_{P_{1}}...\int d\xx_{P_{n}} 
|W(\xx_{P_{1}})|...|W(\xx_{P_{n}})|
\sum_{T}[\prod_{l\in T}
|\b^{-1} g_{\e,\e'}(\xx_l-\yy_l)|]
\Eq(7.32)$$
where we have used that $\int dP_{T}(\tt)=1$.  The number of addends
in $\sum_T$ is bounded by $n!C_2^n$. 

In order to bound the integration over propagators 
we use antiperiodicity 
$$\int_{-\b}^{\b} dx_0 g_{\e,\e'}(\vec x,x_0)
=\int_{-\b/2}^{\b/2} dx_0 g_{\e,\e'}(\vec x,x_0)+
\int_{|x_0|\ge \b/2} dx_0 g_{\e,\e'}(\vec x,x_0)=
\int_{-\b}^{\b} dx_0 (g_{\e,\e'}(\vec x,x_0)+
g_{\e,\e'}(x_0-\b)$$
The tree $T$ realizes a connection
between all the $V$, and we get the bound
$$\int \prod_{i=1}^n d\xx_i {1\over n!}
\sum_{T}[\prod_{l\in T}
|\b^{-1} g_{\e,\e'}(\xx_l-\yy_l)|]\le (\b V) \b^{(n-1)(d+1)}\Eq(bb)$$
In order to perform the integration over the remaining coordinates
we note that if $W(\xx_{P})=V^{-1} \tilde v(x_0)$
then
$$|\tilde v(x_0)|=|{1\over \b}\sum_{k_0\not=0\atop k_0={2\pi\over \b}(n+1)} 
e^{i k_0 t} 
{\k^2\over \k^2+k_0^2}|\le 
{1\over \b}\sum_{n\not=0}  
{\k^2\b^2\over n^2}\le \b^{-1}(\k\b)^2\Eq(F01)$$
so that 
$$\int d\xx V^{-1} |\tilde v(x_0)|\le C 
(\k\b)^2\Eq(f00)$$
if $C$is a suitable constant.
On the other hand if $W(\xx_{P})=W(\xx_1,..,\xx_{2n})$
we use the bound \equ(111.6); then we get, assuming $\k\b\le C^{-1}$
in order to sum over $P_i$
$$|\EE^T(\VV^0;n)|\le n! 
\prod_{i=1}^n[\sum_{P_i} C^{P_i}
|\l|^{max(1,|P_i|/2-1)} (\k^2\b^3)^{max(1,|P_i|/2-1)}]
(\b V) \b^{(n-1)(d+1)}\b^{2 n}\le $$
$$(\b V)
n! C^n \l^n (\k^2\b^3)^{n}  \b^{(d+3)n}\b^{-(d+1)}\le (\b V)
C \l (\k^2\b^{d+6})^n\b^{-(d+1)}n!$$

Finally the following bound can be found, calling 
$\FF_{L,\b,h}=t_{BCS}+\bar\FF_{L,\b,h}$
$$|\bar\FF_{L,\b,h}|\le 
C \l (\k^2\b^{d+6})\b^{-(d+1)}$$
assuming that $\k\le C^{-1}\b^{-d-6\over 2}=\k_0(\b)$ to assuring 
the convergence of the sum over $n$.
\vskip.5cm
{\bf Remark} The above analysis immediately imply a
bound for the {\it effective potential}
$\int P(d\psi^{(i.r.)}) e^{-\VV^0(\psi^{i.r.}+\phi)}$
where $\phi$ is an external fermionic field. The kernels
of the effective potential $W^{(n)}(\xx_1,..,\xx_{n_e})$
at order $n$ obey to the bound
$${1\over V\b}\int d\xx_1..d\xx_n |W_n(\xx_1,..,\xx_n)|\le C^n \l^n
(\k^2\b^{d+6})^n\b^{-{n^e\over 2}}\b^{-(d+1)}\Eq(ricc)$$
\*
as now the propagators are $2n-{n_e\over 2}$
\vskip.5cm
\sub(1.1aklw){\it Extracting a volume factor}

The above analysis says that $\bar\FF_{L,\b,h}$, which is the correction
to the mean field, is given by a convergent expansion for 
sufficiently long range interaction
$0<\k<\k_0(\b)$. We prove now that we can improve the above bound by a factor
$V^{-1}$. 

Consider first the case in which in \equ(tt) there is at
least a $W(P_i)$ associated to $\tilde v$. 
We can write, by using that for the fields 
in $\EE^T$ holds the rule $\psi_\xx=\int d\xx' g(\xx-\xx') {\partial\over\partial\psi_\xx}$
$${1\over n!}{1\over\b V}\EE^T(\VV^0;n)
={1\over n!}{1\over\b V}\{{\l\over V}\int d\xx_1 d\yy_1 d\xx_ad\xx_b d\xx_c d\xx_d\tilde v(x_{0,1}-y_{0,1})$$ 
$$g_{+,\e_a}(\xx_1-\xx_a) g_{+,\e_b}(\xx_1-\xx_b) g_{-,\e_c}(\yy_1-\xx_c)
g_{-,\e_d}(\yy_1-\xx_d)H^{(4,n)}_{\e_a,\e_b,\e_c,\e_d}(\xx_a,\xx_b,\xx_c,\xx_d)$$
$$+2{\l\over V}\int d\xx_1 d\yy_1 d\xx_a d\xx_b g_{+-}(\xx_1-\yy_1)
\tilde v(x_{0,1}-y_{0,1}) g_{+,\e_a}(\xx_1-\xx_a) g_{+,\e_b}(\yy_1-\xx_b)
H^{2,n}_{\e_a,\e_b}(\xx_a,\xx_b)+\Eq(hhf)$$
$$\sum_{\e=\pm}{\l\over V}\int d\xx_1 d\yy_1 d\xx_a d\xx_b g_{-\e,-\e}({\bf 0})
\tilde v(x_{0,1}-y_{0,1}) g_{\e',\e_a}(\xx_1-\xx_a) g_{\e',\e_b}(\yy_1-\xx_b)
H^{2,n}_{\e_a,\e_b}(\xx_a,\xx_b)\}$$
where
$$H^{(4,n)}_{\e_a,\e_b,\e_c,\e_d}(\xx_a,\xx_b,\xx_c,\xx_d)
={\partial\over \partial\psi^{\e_a}_{\xx_a}}
{\partial\over \partial\psi^{\e_b}_{\xx_b}}
{\partial\over \partial\psi^{\e_c}_{\xx_c}}
{\partial\over \partial\psi^{\e_d}_{\xx_d}}\EE^T(\bar V;n-1)\Eq(sd)$$
$$H^{(2,n)}_{\e_a,\e_b}(\xx_a,\xx_b)
={\partial\over \partial\psi^{\e_a}_{\xx_a}}
{\partial\over \partial\psi^{\e_b}_{\xx_b}}\EE^T(\bar V;n-1)\Eq(sd1)$$
Note that the last addend in \equ(hhf) (corresponding to a {\it tadpole}
contribution)
is vanishing; in fact it can be written in momentum space as
$${\l\over V} g(0) {1\over\b} \sum_{p_0\not=0}\d_{p_0,0}\hat v(p_0){1\over \b V}\sum_{\kk'} 
g_{\e',\e_a}(\kk') g_{\e',\e_b}(\kk'+p_0)
H^{2,n}_{\e_a,\e_b}(k',p_0)=0$$
The first addend in \equ(hhf) can be bounded in the following way, remembering that
$|g_{\e,\e'}(\xx,\yy)|\le\b$ 
$$\le C
{\l\over V}\b^{-1}(\k\b)^2 \b^2
\sup_{\xx_a}[\int d\xx_1 
|g_{+,\e_a}(\xx_1-\xx_a)|] \sup_{\xx_c}
[\int d\yy_1 g_{-,\e_c}(\yy_1-\xx_c)]$$
$${1\over n!}{1\over\b V}\int d\xx_a d\xx_b d\xx_c d\xx_d |H^{(4,n)}_{\e_a,\e_b,\e_c,\e_d}(\xx_a,\xx_b,\xx_c,\xx_d)|\Eq(jhf)$$
The bound for the last integral is given by \equ(ricc) with $n^e=4$;
hence the first addend in \equ(hhf) obeys to the following bound
$$C^n {\l^n\over V}\b^{-1}(\k\b)^2 \b^4\b^{2(d+1)}
(\k^2\b^{d+6})^{n-1}\b^{-2}\b^{-(d+1)}
\Eq(ricc1)$$
so that summing over $n$ we have the bound ${\l\over V}(\k\b)^2\b^{d+2}$.

Finally the second addend in \equ(hhf) can be bounded by
$${\l\over V}(\k\b)^2\b \sup_{\xx_a}[\int d\xx_1 
|g_{+,\e_a}(\xx_1-\xx_a)|]\sup_{\xx_b}[\int d\yy_1 
g_{+,\e_b}(\yy_1-\xx_b)]
\int d\xx_a d\xx_b|H^{2,n}_{\e_a,\e_b}(\xx_a,\xx_b)|\Eq(hh)$$
and again using 
\equ(ricc) with $n^e=2$ we get that
hence the second addend in \equ(hhf) obeys to the following bound
$$C^n {\l^n\over V}\b^{-1}(\k\b)^2 \b^3\b^{2(d+1)}
(\k^2\b^{d+6})^{n-1}\b^{-1}\b^{-(d+1)}
\Eq(ricc2)$$
Of course there is no $\tilde v$, we can apply the same reasoning to 
one of the kernel $W(\xx_1,..\xx_n)$.
Summing over $n$ we have the bound ${\l\over V}(\k^2\b^3)\b^{d+2}$; then, for $k\le k_0$ we get
the better bound
$$|\bar\FF_{L,\b,h}|\le C {\l\over V}(\k^2\b^3)\b^{d+2}\Eq(tty)$$
\vskip.5cm
\*
\sub(1.1akl){\it The integration of $S$}

By performing the change of variables, if
$\psi=(\psi^+,\psi^-)$ and $g$ is the matrix propagator
of $P_\s(d\psi)$,
$\psi_\kk\to \psi_\kk+g \psi_\kk$, 
we get for two point
Schwinger function the formula
$$S_{\e,\e'}(\kk,u,v)
=g_{\e,\e'}(\kk)+\sum_{\e'',\e'''} g_{\e,\e''}(\kk) V_{2;\e',\e'''}(\kk)
g_{\e''',\e'}(\kk)\Eq(mm)$$
where $V_{2;\e',\e'''}(\kk)$ is the kernel of the effective potential 
with two external fields; it can be bounded by
$$|V_{2;\e,\e'}(\kk)|\le {1\over \b V}\int d\xx 
\int d\yy| V_{2;\e,\e'}(\xx,\yy)|\Eq(kbda)$$
By using \equ(ricc) we get that  
$$|V_{2;\e,\e'}(\kk)|\le C^n \l^n
(\k^2\b^{d+6})^n\b^{-1}\b^{-(d+1)}\Eq(rfr)$$
By can improve the above bound as described in \S 3.3.
If there is at least a $W=V^{-1}\tilde v$, we get
$$V_{2;\e,\e'}(\xx,\yy)
={\l\over V}\int d\zz d\zz' 
d\xx_1d\xx_2d\xx_3d\xx_4
g(\zz-\xx_1)
g(\zz-\xx_2) g(\zz'-\xx_3)
g(\zz-\xx_4) V_6(\xx,\yy,\xx_1,\xx_2,\xx_3,\xx_4)+$$
$${\l\over V}\int d\xx 
\int d\yy d\xx_1 d\xx_2 \int d\zz d\zz' g(\zz-\zz')
g(\zz-\xx_1) g(\zz'-\xx_2)
V_4(\xx,\yy,\xx_1,\xx_2)$$
where we have used that the tadpoles contributions is vanishing.
The the integral over $\xx,\yy$ times $(\b V)^{-1}$ of the first addend can 
be bounded, using also that $sup |g|\le\b$ 
$$
{\l\over \b V}(\k\b)^2{1\over \b V}\int d\xx 
\int d\yy d\xx_1d\xx_2d\xx_3d\xx_4\int d\zz d\zz' $$
$$|g(\zz-\xx_1)g(\zz-\xx_2) g(\zz'-\xx_3)
g(\zz-\xx_4)| |S(\xx,\yy,\xx_1,\xx_2,\xx_3,\xx_4)|\le $$
$${\l\over \b V}(\k\b)^2\b^2 (sup_{\xx_1}|\int d\zz g(\zz-\xx_1)|)
(sup_{\xx_3}|\int d\zz' g(\zz'-\xx_3)|)
{1\over \b V}\int  d\xx 
\int d\yy d\xx_1d\xx_2d\xx_3d\xx_4 |S(\xx,\yy,\xx_1,\xx_2,\xx_3,\xx_4)|\Eq(mm)$$
By using \equ(ricc) we get for \equ(mm) the bound
${\l\over V}(\k\b)^2 \b^{d+1}$.
On the other hand the second addend is bounded by
$${1\over \b V^2}\b (sup_{\xx_1}|\int d\zz g(\zz-\xx_1)|)
(sup_{\xx_3}|\int d\zz' g(\zz'-\xx_3)|)
\int  d\xx 
\int d\yy d\xx_1d\xx_2 |V_2(\xx,\yy,\xx_1,\xx_2)|$$
from again by using
 \equ(ricc) we get the same bound ${\l\over V}(\k\b)^2 \b^{d+1}$.
If there are no vertex $\tilde v$, we can repaet the above argument on the 
kernels of \equ(111.6).
Hence from \equ(mm)
$$|S_{\e,\e'}(\kk,u,v)
-g_{\e,\e'}(\kk)|\le 
{\l\over V}(\k^2\b^3) \b^{d+1}\Eq(gg1)$$
%
%
%
%
%
%

\appendix (A1, The ultraviolet integration)

The integration of the ultraviolet part \equ(fs1)
can be done by a multiscale analyses; it is quite standard
and we refer to \S 3 of [BM] for details in a similar case.
It is convenient to
introduce an ultraviolet cut-off $N$ by writing 
$$g^{[1,N]}(\xx)=\sum_{k=1}^N g^{(k)}(\xx)\Eq(A1.1)$$
where
$$g^{(k)}(\xx)={1\over V\b}\sum_{\kk\in\DD_{\b,L}} 
h_k(k_0)\fra{e^{-i\kk\xx}}{-ik_0+\e(\vec k)-\m}\Eq(A1.2)$$
with $h_k(k_0)=H_0(\g^{-k}|k_0|)-
H_0(\g^{-k+1}|k_0|)$. 
Note that
$\lim_{N\to\io} g^{[1,N]}(\xx)=g^{(u.v.)}(\xx)$ and that,
for any integer $K\ge 0$, 
$g^{(k)}(\xx)$ satisfies the bound, for any integer $K$
$$|g^{(k)}(\xx)|\le {C_K\over 1+(\g^k|x_0|+|\vec x|)^K}\Eq(A1.3)$$
We associate to any propagator $g^{(k)}(\xx)$ a Grassmann field 
$\psi^{(k)}$ and a Gaussian integration $P(d\psi^{(k)})$ with 
propagator $g^{(k)}(\xx)$. We can rewrite $\VV^{(0)}$ as:
$$\VV^{(0)}(\phi)+V\b E_1=-\lim_{N\to\io}\log 
\int P(d\psi^{(0)})P(d\psi^{(1)})\cdots 
P(d\psi^{(N)}) e^{-V(\psi^{[1,N]}+\phi) }\Eq(A1.4)$$
We can integrate iteratively the fields on scale $N,N-1,\ldots,k+1$ and 
after each integration, using iteratively an identity like \equ(ass),
we can rewrite the r.h.s. of 
\equ(A1.4) in terms of a new effective potential $\VV^{[1,k]}$:
$$\equ(A1.4)=\lim_{N\to\io}\Big\{
V\b\sum_{j=k+1}^N E_j-\log 
\int P(d\psi^{(1,0)})P(d\psi^{(1,1)})\cdots 
P(d\psi^{(1,k)}) e^{-\VV^{(k,N)}(\psi^{[1,k]}+\phi) }\Big\}\Eq(A1.5)$$
with $\VV^{(k,N)}(\psi^{[1,k]})$ admitting a representation 
in terms of {\it trees} defined in the following way:

\0 1) Let us consider the family of all trees which can be constructed
by joining a point $r$, the {\it root}, with an ordered set of $n\ge 1$
points, the {\it endpoints} of the {\it unlabeled tree}, 
so that $r$ is not a branching point. $n$ will be called the
{\it order} of the unlabeled tree and the branching points will be called
the {\it non trivial vertices}.
The unlabeled trees are partially ordered from the root to the endpoints in
the natural way; we shall use the symbol $<$ to denote the partial order.
 
Two unlabeled trees are identified if they can be superposed by a suitable
continuous deformation, so that the endpoints with the same index coincide.
It is then easy to see that the number of unlabeled trees with $n$ end-points
is bounded by $4^n$.
 
We shall consider also the {\it labeled trees} (to be called simply trees in
the following); they are defined by associating some labels with the unlabeled
trees, as explained in the following items.
 
\0 2) We associate a label $k\ge 0$ with the root and we denote 
$\TT_{(k,N),n}$ the
corresponding set of labeled trees with $n$ endpoints. Moreover, we introduce
a family of vertical lines, labeled by an an integer taking values in
$[k,N]$, 
and we represent any tree $\t\in\TT_{(k,N),n}$ so that, if $v$ is an
endpoint or a non trivial vertex, it is contained in a vertical line with
index $h_v>k$, to be called the {\it scale} of $v$, while the root is on the
line with index $k$. 
There is the constraint that, if $v$ is an endpoint, $h_v=N+1$.
 
The tree will intersect in general the vertical lines in set of
points different from the root, the endpoints and the non trivial vertices;
these points will be called {\it trivial vertices}. The set of the {\it
vertices} of $\t$ will be the union of the endpoints, the trivial vertices
and the non trivial vertices.
Note that, if $v_1$ and $v_2$ are two vertices and $v_1<v_2$, then
$h_{v_1}<h_{v_2}$.
 
Moreover, there is only one vertex immediately following
the root, which will be denoted $v_0$ and can not be an endpoint;
its scale is $k+1$.

\0 3) With each endpoint $v$ of scale $h_v=N+1$ we associate 
$\hat\VV$ \equ(p14).
Given a vertex $v$, which is not an endpoint, $\xx_v$ will denote the family
of all space-time points associated with one 
of the endpoints following $v$.  

\0 4) The trees containing only the root and an endpoint of scale $k+1$ will
be called the {\it trivial trees}.

\0 5) We introduce a {\it field label} $f$ to distinguish 
the field variables
appearing in the terms $\tilde\VV$ 
associated with the endpoints. 
The set of field labels associated with the endpoint $v$ will be called 
$I_v$.
Analogously, if $v$ is not an endpoint, we shall
call $I_v$ the set of field 
labels associated with the endpoints following
the vertex $v$; $\xx(f)$, and $\s(f)$ will denote the space-time
point, the $\s$ index and the $\o$ index, respectively, of the
field variable with label $f$.

The effective potential can be written in the following way:
$$\VV^{(h)}(\psi^{(\le h)}) + L\b \tilde E_{h+1}=
\sum_{n=1}^\io\sum_{\t\in\TT_{(h,N)n}}
V^{(h)}(\t,\psi^{(\le h)})\Eq(3.27)\;,$$
where, if $v_0$ is the first vertex of $\t$ and $\t_1,..,\t_s$ ($s=s_{v_0}$)
are the subtrees of $\t$ with root $v_0$,\\
$V^{(h)}(\t,\psi^{(\le h)})$ is defined inductively by the relation
$$\eqalign{
&\qquad V^{(h)}(\t,\psi^{(\le h)})=\cr
&{(-1)^{s+1}\over s!} \EE^T_{h+1}[
V^{(h+1)}(\t_1,\psi^{(\le h+1)});..; 
V^{(h+1)}(\t_{s},\psi^{(\le h+1)})]\;,\cr}\Eq(3.28)$$
and $\bar V^{(h+1)}(\t_i,\psi^{(\le h+1)})$,
if $\t_i$ is trivial and $h=N-1$, it
is equal to $\hat\VV$ \equ(p13).

By iterating \equ(3.28) we can write 
$V^{(h)}(\t,\psi^{(\le h)})$ in the following way.
We associate with any vertex $v$ of the
tree a subset $P_v$ of $I_v$, the {\it external fields} of $v$.
These subsets must satisfy various constraints. First of all, if $v$ is not
an endpoint and $v_1,\ldots,v_{s_v}$ are the vertices immediately following it,
then $P_v \subset \cup_i P_{v_i}$; if $v$ is an endpoint, $P_v=I_v$. We shall
denote $Q_{v_i}$ the intersection of $P_v$ and $P_{v_i}$; this definition
implies that $P_v=\cup_i Q_{v_i}$. The subsets $P_{v_i}\bs Q_{v_i}$,
whose union will be made, by definition, of the {\it internal fields} of $v$,
have to be non empty, if $s_v>1$.
 
Given $\t\in\TT_{(h,N)n}$, there are many possible choices of the subsets $P_v$,
$v\in\t$, compatible with all the constraints; we shall denote $\PP_\t$ the
family of all these choices and $\bP$ the elements of $\PP_\t$. Then we can
write
$$V^{(h)}(\t,\psi^{(\le h)})=\sum_{\bP\in\PP_\t}V^{(h)}(\t,\bP)
\;;\Eq(3.36)$$
$V^{(h)}(\t,\bP)$ can be represented as 
$$V^{(h)}(\t,\bP)=\int d\xx_{v_0} \tilde\psi^{(\le h)}
(P_{v_0}) K_{\t,\bP}^{(h+1)}(\xx_{v_0})\;,\Eq(3.37)$$
with $K_{\t,\bP}^{(h+1)}(\xx_{v_0})$ defined inductively (recall that $h_{v_0}
=h+1$) by the equation, valid for any $v\in\t$ which is not an endpoint,
$$\eqalign{
K_{\t,\bP}^{(h_v)}(\xx_v)&={1\over s_v !}
\prod_{i=1}^{s_v} [K^{(h_v+1)}_{v_i}(\xx_{v_i})]\;\cdot\cr
&\cdot\;\EE^T_{h_v}[ \tilde\psi^{(h_v)}(P_{v_1}\bs Q_{v_1}),\ldots,
\tilde\psi^{(h_v)}(P_{v_{s_v}}\bs Q_{v_{s_v}})]\;,\cr}\Eq(3.38)$$
where iv $v$ is an endpoint 
$K^{(h_v)}_{v}(\xx_v)$ is the kernel $\tilde v(\xx)$.
We call  $\c$--vertices the vertices $v$ of $\t$ such that 
$\EE^T_{h_v}$ is not trivial.

By using the representation 
of the truncated expectation analogous to
\equ(7.19) and the Gram inequality we get that
the contribution from a tree $\t\in\TT_{[1,k],n}$
associated to a kernel with $2l$ external legs can be bounded as
(see \S 3.14 [BM] for details in a similar case):
$$\eqalign{&\fra{1}{V\b}\int d\xx_1\cdots d\xx_{2l}
|W^{(k,N)}_{2l}(\t;\xx_1,\s_1,\e_1;\ldots;
\xx_{2l},\s_{2l},\e_{2l})| \le\cr
&\hskip2.truecm \le C^n |\l(\k\b)^2|^n \g^{-k(n-1)}
\prod_{v\ {\rm not}\ {\rm e.p.}}\g^{-(h_v-h_{v'})
(n_v-1)}\;,\cr}\Eq(A1.6)$$ 
where $v'$ is the vertex immediately preceding $v$ on $\t$, 
$n_v$ is the number
of endpoints following $v$ on $\t$.

Note that the "dimension" $n_v-1$ is vanishing when 
when $n_v=1$, so that the above bound is not suitable to sum
over trees. However in the case $n_v=1$ 
the above 
bound can be improved. If fact let be $v$ a vertex with $n_v=1$
such that its set of internal lines is not empty; to such vertex 
is then associated a truncated expectation
$$\EE^T_{h_v}(\hat\VV)=-{\l\over V} \sum_{\s,\s'}
\int d\xx\int d\yy \tilde v(x_0-y_0)
\psi^+_{\xx,\s}\psi^-_{\yy,\s'} g^{h_v}_{+-}(\xx,\yy)\Eq(td)$$
where we have used that the contration of $\psi^+\psi^+$
or $\psi^-\psi^-$ in $\hat\VV$ is vanishing by momentum conservation
(remember that $p_0=0$). In \equ(A1.6)
$g^{h_v}_{+-}(\xx,\yy)$ is bounded with a constant;
however such bound can be improved by writing
$$g_{+-}^{h_v}(\xx;\yy)=g_{+-}^{h_v}({\vec x},x_0;{\vec y},x_0)+
(x_0-y_0)\int_0^1 
dt\partial_{x_0} g^{h_v}_{+-}
({\vec x},x_0;{\vec y},y_0+t(x_0-y_0)))\Eq(int)$$
and noting that 
$$|g^{h_v}_{+-}(\xx,\xx)|\le |{1\over V\b}
\sum_\kk e^{-i{\vec k}{\vec x}}f_k(k_0){-i k_0\over k_0^2+E^2(\vec k)+|\s|^2}|
+|{1\over V\b}
\sum_\kk e^{-i\vec k{\vec x}}f_k(k_0){E(\vec k)\over k_0^2+E^2(\vec k)+|\s|^2}|\le 
C\g^{-h_v}\Eq(bou)$$
Then the contribution to the kernel
of \equ(td) coming from the first addend in the r.h.s.
of \equ(int) is bounded by $C\g^{-h_v}$; in the contribution
coming from the second addend we bound the propagator by
$C\g^{-h_v}$ so getting a factor $V^{-1}\int d{\bf r}
|r_0||\tilde v(r_0)|\le C\b$, so that at the hand we get
a bound for \equ(td) $C\b\g^{-h_v}$.

Then we get the bound
$$\eqalign{&\fra{1}{V\b}\int d\xx_1\cdots d\xx_{2l}
|W^{(k,N)}_{2l}(\t;\xx_1,\s_1,\e_1;\ldots;
\xx_{2l},\s_{2l},\e_{2l})| \le\cr
&\hskip2.truecm \le C^n |\l(\k\b)^2|^n \g^{-k(n-1+n^{tad})}
\b^{n}
\prod_{v\ {\rm not}\ {\rm e.p.}}\g^{-(h_v-h_{v'})
(n_v-1+z_v)}\;,\cr}\Eq(A1.6)$$ 
\\
where $z_v=1$ if $n_v=1$ and $0$ otherwise, and $n^{tad}$
is the total number of $c$-vertices $v$ with $n_v=1$ 
such that its set of internal lines is not empty.
Then, proceeding as in \S 3.14 of [BM], one can sum over $\t$
and the bound \equ(111.6) is proved.

Finally by proceedig as in \S 3.3 it is easy to see that we can extract 
a factor $O(V^{-1})$ from each kernel $W$.

\vskip1cm

\centerline{\titolo References}

\*

\halign{\hbox to 1.2truecm {[#]\hss} &
        \vtop{\advance\hsize by -1.25 truecm \0#}\cr

BCS& {Bardeen J., Cooper L.N.,Schreifer J.R. {\it Phys. Rev.}
108,1175 (1957). }\cr
B& {Bogolubov, N.N. {\it J.E.T.P.} 7,41 (1958).}\cr
BM&{ Benfatto, G, Mastropietro, V. {\it Rev. Math. Phys.} 13,
11,1323--1425 (2001).}\cr
BR&{Bardeen,J,Rickayzen, G. {\it Phys. Rev.} 118,936 (1960).}\cr
%
%
BZT&{Bogolubov,N.N., Zubarev D.N.,Tserkovnikov I.A. {\it Sov. Phys.-Doklady}
2,535 (1958)}\cr
CEKO&{Carlson E.W.,Emery V.J.,Kivelson S.A.,Orgad D. cond-mat0206217.}\cr
GLM&{Gallavotti, G, Lebowitz L, Mastropietro V. {\it J. Stat. Phys.}
108, 5, 831--861 (2002).}\cr
GM& {G. Gentile, V. Mastropietro.{\it Phys. Rep.} {\bf
352}, 273--437 (2001). }\cr
H& {Haag, R. {\it Nuovo Cimento} 25,287 (1962).}\cr
L& {Lehmann D. {\it Comm. Math. Phys.} 198,427--468 (1998).}\cr
LMP& {Lebowitz J, Presutti,E, Mazel. {\it Phys. Rev. Lett.}}\cr 
M& {Muhlschlegel B. {\it J. Math. Phys.} 3, 522--530 (1962).}\cr
SHML& {Salmhofer, M.,Honerkamp C, Mtzner W, Lauscher O. cond-mat 0409725}\cr
T& {Thirring W. {\it Comm. Math. Phys.} 7,181--189 (1967).}\cr
TW& {Thirring W., Wehrl W.
{\it Comm. Math. Phys.} 4,303--314 (1966).}\cr
}

\bye